# *TasVisAn* and *InsPy* – Python Packages for Triple-Axis Spectrometer Data Visualization, Analysis, Instrument Resolution Calculation, and Convolution

**Guochu Deng[1]*, Garry J. McIntyre[1]**

[1]Australian Centre for Neutron Scattering, Australian Nuclear Science and Technology Organisation, New Illawarra Road, Lucas Heights, NSW 2234, Australia

**Synopsis** *TasVisAn* and *InsPy* are two Python packages for data reduction, visualization, and analysis, instrument resolution calculation, and resolution convolution fitting for both traditional single-detector and modern multi-analyser triple-axis spectrometers.

**Abstract**

Experimental data collected from a triple-axis spectrometer (TAS) are typically analysed by considering the instrument resolution, as the resolution of a TAS instrument is often complex and significantly influences the measured results. Two Python packages, *TasVisAn* and *InsPy*, have been developed to visualize and analyse data from TAS instruments—particularly from the cold-neutron TAS Sika and the thermal-neutron TAS Taipan at the Australian Centre for Neutron Scattering. *TasVisAn* offers a range of functions, including data importing, reduction, plotting, contour mapping, convolution fitting, and more, for data collected on TAS instruments, especially on Sika and Taipan. It also supports data reduction of the current trendy multi-analyser and multiplexing TAS instruments, including the multiplexing mode of Sika. Besides, it includes scan simulation and batch file validation tools for both Taipan and Sika, assisting users in designing and planning experiments in advance. *InsPy* is a general-purpose Python package designed to calculate the four-dimensional (4D) instrument resolution in momentum–energy space for any TAS instrument. Combined with *InsPy*, *TasVisAn* supports both instrument resolution calculation and resolution-convoluted data fitting. Its flexible external data import feature further allows *TasVisAn* to be adapted for the visualization and convolution analysis of inelastic neutron scattering data across various TAS instruments.

**Keywords:** Triple-Axis Spectrometer, Resolution Function, Data Reduction, Data Visualization, Convolution Fitting

*Corresponding Author, Email: guochu.deng@ansto.gov.au

## 1. Introduction

A triple-axis spectrometer (TAS) instrument is a widely used neutron instrument for inelastic neutron scattering experiments. It was first invented by Dr. Bertram N. Brockhouse at the Chalk River Reactor in Canada, who won the Noble prize for this invention and his contribution in magnetism.(Brockhouse, 1995) A TAS instrument includes three main axes: the monochromator axis, the sample axis, and the analyser axis. The monochromator and analyser of a TAS instrument are made up of arrays of crystals, which are mounted on a mechanical device with the capability to implement the vertical and horizontal curvings to focus neutron beam at the sample position.(Shirane *et al.*, 2004)

Considering the relatively lower flux of neutron instruments comparing to X-ray instruments at a synchrotron beamline, crystals with large mosaicity such hot-pressed pyrolytic graphite (PG) are adopted to fabricate the monochromator and analyser of a TAS instrument. Thus, higher neutron flux can be achieved at the sample position, critical for collecting weak inelastic neutron scattering signals. However, the trade-off of the high flux strategy is that the instrument resolution is deteriorated, especially in energy scale. The instrumental resolution of a TAS instrument is a four-dimensional (4D) ellipsoid in the momentum-energy space.(Shirane *et al.*, 2004, Zheludev, 2009) Due to different instrument geometries, crystal mosaicities, and other configurations, TAS instrument resolution varies significantly in the different combinations of the momentum and energy. This causes the complexity to analyse the experimental data from a TAS instrument. A program, which can fit experimental data with convolution of the instrument resolution, is indispensable to accurately analyse TAS data. *Reslib* was developed in MATLAB(The MathWorks, 2023) for this purpose and was widely used in by TAS users.(Zheludev, 2009). Several other software, such as *ResLibCal* (Farhi & Willendrup, 2018)and *RESTRAX*,(Šaroun & Kulda, 1997) were developed for the similar purpose as well.

In recent years, Python,(Foundation, 2024) as a popular programming language, is widely used in many different fields, especially, in the areas of data science and artificial intelligence. First of all, comparing to other commercial software like MATLAB,(The MathWorks, 2023) OriginLab,(Corporation, 2022) and IDL,(L3Harris Geospatial Solutions, 2022) Python is completely free. Users of neutron and synchrotron radiation facilities come from different countries and background. Free software provides them the equivalent opportunities to use the software without financial barriers, regardless of their institution's funding level or geographic location. Second, many open-source packages such as the array operation package *numpy*,(Harris *et al.*, 2020) the table-based data manipulation and analysis package *pandas*,(McKinney, 2010) the data visualization packages *matplotlib(Hunter, 2007)* and *plotly (Inc., 2015)* are available and well maintained. Furthermore, combining with another powerful note-taking Python package, *Jupyter notebook*,(Thomas Kluyver *et al.*, 2016) a Python-based software, providing an alternative solution to MATLAB, OriginLab or IDL, can not only help to users to quickly plotting and fitting their experimental data, but also assist them to quickly tidy up their experimental log. With these advantages, Python becomes a first choice for data analysis at large facilities such as neutron and synchrotron facilities. For example, The European spallation source (ESS) has adopted Python as their primary language for the data reduction and analysis.(McCluskey *et al.*, 2024) Many neutron-scattering software are based on Python as well. For example, the Mantid project is a large Python-based neutron data analysis platform.(Arnold *et al.*, 2014) It becomes more and more widely accepted by neutron users from all over the world as a standard neutron data analysis tool. Thus, a package written in Python will be easier to communicate and integrate with Mantid for its specific functions, which could greatly enhance the data analysis efficiency of users. One attempt to implement a comprehensive Python package for data import and resolution calculation of TAS instruments was *neutronpy* created by David M Fobes.(Fobes, 2018) However, this project has not been updated since 2019 and some planned functions have not been implemented yet.

In this work, we present the two Python packages, *TasVisAn* and *InsPy*, for the data analysis and the resolution calculations of TAS instruments. The structures and functions of these two packages will be explained in detail. Some examples will be introduced to demonstrate various functions provided by these Python packages.

## 2. Design and Implementation

To promote modularity, reusability, and maintainability, both the *TasVisAn* and *InsPy* packages are structured using object-oriented programming principles. Scientific data analysis often involves complex objects such as experiments, datasets, and instruments, each with associated operations. Organizing these as classes allows for a natural mapping between scientific concepts and code structure, facilitating both user understanding and future extension of the software.

*TasVisAn* consists of a parent class called TasData to represent general traditional TAS data which involves the standard data columns such as "*qh*", "*qk*", "*ql*", "*en*", "*m1*", "*m2*", "*s1*", "*s2*", "*a1*", "*a2*", "*detector*", "*monitor*", etc. This parent class implements all the data manipulation, such as combination, normalization, correction, etc. and data visualization, including various line plots and contour map. It has an experiment variable as its property to keep the TAS instrument configuration, which can be used for experiment simulations and resolution calculations. TasData class is not only designed for traditional single-detector TASs, but also designed to treat data from multi-analyser TASs, which are the current trend of TAS instruments. *TasVisAn* also implements two main classes, Sika and Taipan, which are inherited from the parent class TasData**.** Sika and Taipan implement the data import and data reduction from the data collected on the cold-neutron TAS Sika(Wu *et al.*, 2016, Deng & McIntyre, 2020) and the thermal-neutron TAS Taipan(Danilkin & Yethiraj, 2009, Rule *et al.*, 2018) at the Australian Centre for Neutron Scattering (ACNS), respectively. Considering that Sika and Taipan have different names for their mechanical motors and virtual motors, the data reduction function will convert all these mechanical and virtual motor names into the standard names in TasData. Thus, both Sika and Taipan classes share the same functions, which is convenient for users to use. This feature substantially enhances the ease of use of this package. Such a design also greatly improves the scalability and flexibility of the package. For example, *TasVisAn* can be easily extended to another TAS instrument by simply creating a new instrument subclass inheriting from TasData with its own implementation of the data reading. Namely, the only necessary code for this new instrument is the function to import the original instrument data file and convert to the standard format of TasData. The current structure design allows us to extend *TasVisAn* to cover more TAS instruments without extensive efforts. This will significantly benefit users who need to analyse data from other facilities or combine data from different instruments cross platforms.

As mentioned above, the instrumental resolution convolution calculation is indispensable for the TAS data analysis. To implement this function in *TasVisAn*, we developed another Python package *InsPy* to calculate TAS instrument resolutions, following the methodology of the MATLAB package *Reslib* and partially derived from the Python package *neutronpy*. *InsPy* not only extends the capabilities of *neutronpy* but also improves its code stability and resolves issues present in the original implementation, leading to more reliable performance in practical use. *InsPy* implement a main class *TripleAxisSpectr* to hold the instrument configuration and conduct the resolution calculation. Upon inputting the instrument configuration, *TripleAxisSpectr* can easily calculate the instrument resolution at any given $Q$ and energy in the 3D $Q_x$-$Q_y$-$E$ space. The projected resolution ellipses are calculated at the same time. With the calculated resolution function, one can fit the TAS data by convoluting the resolution using the *FitConv* class in *InsPy*.

Combining *TasVisAn* and *InsPy*, one can conduct full data analysis such as data reduction, combination, normalization, visualization, and the convolution fitting. With the scan simulation function provided by TasVisAn, users can simulate possible scans on Taipan and Sika to plan their experiments in detail before their experiments start. They could even start to write their batch files and validate the commands in their batch files to avoid any mistakes in the experiment. During their experiments or after their experiments, users could export the experiment logfiles including the scan numbers, scan commands, and the corresponding scan titles for the convenience of taking notes for their experiments and recording their discoveries during experiments. *TasVisAn* and *InsPy* are primarily designed for the post-experiment data analysis. With the instrument configurations as inputs, users can conduct batch fitting to their data with a few lines of Python code and generate publishable figures without pain. Namely, *TasVisAn* and *InsPy* provide a full set of comprehensive functions to users for planning and running experiments, taking experiment notes, data analysis, and plotting figures, which significantly improve users' experimental experiences on Sika and Taipan and their work efficiency. Table 1 lists the main classes and most important functions in the two packages.

Both the source codes of TasVisAn and Inspy are available online. *TasVisAn* is available for download at https://github.com/gcdengansto/TasVisAn and *Inspy* is available for download at https://github.com/gcdengansto/inspy. Considering some basic knowledge of Python scripting is necessary to use the full potential of the packages *TasVisAn* and *InsPy*, an example of Jupyter notebook is included in the source code package of TasVisAn, which can help users quickly start to use these two packages without great efforts. Additionally, several user-friendly Graphical User Interfaces (GUIs) are available to assist users without programming skills in browsing and fitting data. The demonstration videos about how to use these GUIs can be found in the supporting information.

Table 1. The main classes and functions in the *TasVisAn* and *Inspy* packages

| Package | Main Class | | Main functions |
|---|---|---|---|
| *TasVisAn* | TasData (parent) | | `tas_datacombine`<br>`tas_datanormalize`<br>`tas_combplot`<br>`tas_export_hklw`<br>`tas_tidy_contour`<br>`tas_expconfig`<br>`tas_convfit`<br>`tas_multichannel_config`<br>`tas_multichannel_reduction`<br>`tas_multichannel_plot3d`<br>... |
| | | Taipan(child) | `taipan_data_to_pd` |
| | | | taipan_hdf_to_pd<br>taipan_scanlist_to_dflist<br>taipan_batch_reduction<br>taipan_reduction_by_row<br>taipan_reduction_anycolumn<br>taipan_export_hklw<br>taipan_calibr_6scans<br>calibr_offset_from_fit<br>taipan_batch_validate<br>taipan_scansim<br>taipan_convfit<br>... |
| | | Sika(child) | sika_data_to_pd<br>sika_hdf_to_pd<br>sika_scanlist_to_dflist<br>sika_batch_reduction<br>sika_reduction_by_row<br>sika_reduction_anycolumn<br>sika_export_hklw<br>sika_batch_validate<br>sika_scansim<br>sika_convfit<br>sika_multichannel_config<br>sika_multichannel_reduction<br>sika_multichannel_plot3d<br>... |
| | | Other TASs (e.g., BT7) | (To be developed) |
| | TaipanCommandValidator | | validate_command |
| | SikaCommandValidator | | validate_command |
| | Data Browser GUI | | Integrated GUI for Taipan & Sika |
| *Inspy* | TripleAxisSpectr | | CalcResMatQ<br>CalcResMatHKL<br>ResConv<br>ResConSMA<br>ResolutionPlot<br>ResolutionPlotProj<br>ResolutionPlot3D<br>... |
| | FitConv | | Fitwithconv<br>... |
| | Resolution Calculation & Data Fitting GUI | | GUI for Escan, GUI for Qscan |
| | TimeOfFlightSpectr | | (To be developed) |

### 3. Features and Capabilities

#### 3.1 *TasVisAn* Package

*TasVisAn* is a Python package aiming to provide a set of handy tools for TAS users to efficiently reduce, combine, normalize, plot, and analyse their data collected on a TAS instrument. It can also help users to plan and simulate their experiments precisely in advance, and to quickly track their data collection and validate their batch files during experiments. In the following, the most important functions will be introduced in detail.

#### 3.1.1 TAS Instrument Calibration

A TAS instrument needs calibration to correct its incident neutron energy and the zero position of the sample scattering angle $2\theta$ (denoted as *s2* in the context of a TAS instrument) from time to time. In order to calibrate these motor values, a powder sample such as a Ni powder sample can be used for the calibration. A series of diffraction peaks of Ni powder, including some high-order lambda-half peaks, which are widely and roughly evenly distributed within the instrument scattering angle limit, will be chosen to be measured in the so-called two-axis mode of the TAS instrument. Namely, the analyser and detector angles of the TAS instrument will be set to zero to create a straight detector geometry. The analyser will be moved to an offset position to allow neutrons to fly directly to the detector. In this setup, the nickel powder diffraction pattern is collected. The two parameters which determines the peak positions are the wavelength and the offset of the sample scattering angle $\Delta_{s2}$.

According to the Bragg law:

$$\lambda = 2d_{Ni(hkl)}\sin\left(\frac{P_{Ni(hkl)}-\Delta_{s2}}{2}\right) \tag{1}$$

where λ is the incident neutron wavelength, $d_{Ni(hkl)}$ are the lattice constants of (*hkl*) reflections of the Ni powder, $P_{Ni(hkl)}$ the peak positions of the *(hkl)* reflections of the Ni powder and $\Delta_{s2}$ is the offset angle of the sample scattering angle *s2* of the instrument. On the other hand, the monochromator (assuming using PG(002) crystals) angle determines the wavelength of the incident neutron beam:

$$\lambda = 2d_{PG(002)}\sin\left(\frac{m2}{2}\right) \tag{2}$$

where $d_{PG(002)}$ is the lattice constant of PG (002) reflection. *m2* is the monochromator scattering angle.

Joining the two equations above, the new equation links *m2* and $\Delta_{s2}$ together with the known values of $d_{Ni(hkl)}$ and $d_{PG(002)}$) and the measured $P_{Ni(hkl)}$ positions. By collecting several peak positions, a nonlinear fit of these peak positions to the equation above could help us to obtain the optimized values for the two parameters, namely, precisely determining the values of *m2* and $\Delta_{s2}$. Thus, the instrumental calibration can be accomplished.

In the Taipan class, two functions are provided to quickly calibrate the instrument. The function `taipan_calibr_6scans` can fit six Ni peaks using the Python package *lmfit(Newville et al., 2016)* and generate the fitted peak positions as shown in Figure1. These peak positions will be used for the function `calibr_offset_from_fit` to conduct the least square fitting and obtain the current position of *m2* and the offset of *s2*, which can be used to correct the motor positions of these two angles. Following this step, the analyser can be calibrated by conducting multiple *a1* and *a2* scans and their coupled scans.

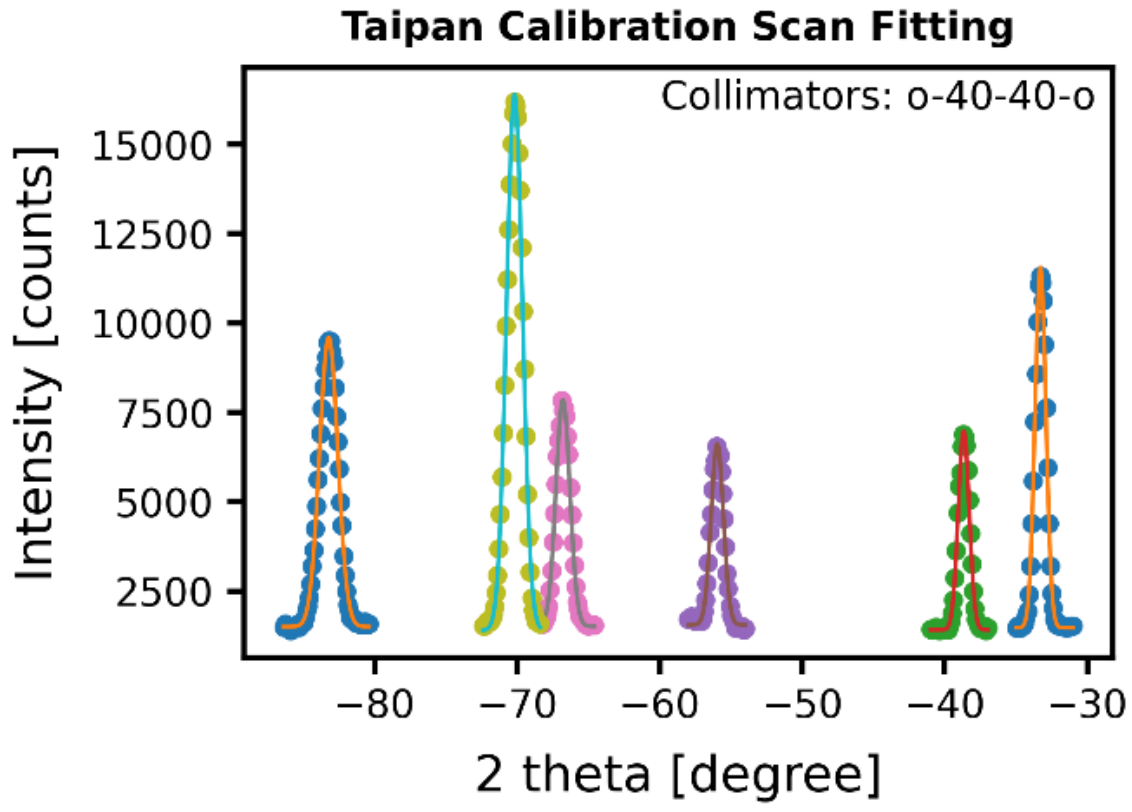

Figure 1 Fitted Ni powder diffraction peaks measured for calibrating Taipan with $E_i$=14.87meV and the open-40-40-open collimation.

### 3.1.2 Simple TAS Data Reduction

In a TAS experiment, inelastic scans normally involved the *qh, qk,* and *ql* positions of the single crystal sample and the energy transfer *en* of the instrument. These data are essential for the data fitting and plotting. However, the original data contain information of the real motor positions. Thus, data reduction is an essential first step for the data visualization and analysis. *TasVisAn* provides the data reduction function in both Taipan and Sika classes. The function `taipan_batch_reduction` and `sika_batch_reduction` could extract the essential *qh, qk, ql*, and *en* positions and the counts from the detector for data plotting and fitting. The extracted data can be saved to a *hklw* text file with a columnar structure.

Another function `reduction_by_row` is designed for any type of data reduction on Taipan and Sika. Users can choose the columns they are interested in for the data reduction. For example, they may be interested in the temperature-dependent or magnetic-field-dependent of peak intensities. Using `reduction_by_row`, they can easily extract the temperature or magnetic field values together with the peak intensities for plotting or any other purposes. This function will be demonstrated in the example section.

Since it is normal operations to repeat scans to improve data quality and add extra data points to the obtained scans to improve the peak features with further details in neutron experiments, the functions such as data combination, normalization, and manipulation are highly important for data analysis. In both classes, it is easy to combine these scans simply by putting the scan numbers into one set of square brackets such as [12335, 12365]. For example, `taipan_batch_reduction` ([12316, 12317, [12318, 123526], 12319, 12320]), which means that the two scans [12318, 123526] in the inner bracket will be combined into one scan. Normalization is usually automatically implemented as the default setting of most of the functions. With normalized data, users can easily manipulate their data, such as scan additions and background subtractions.

### 3.1.3 Multi-analyser TAS Data Reduction

In recent years, multi-analyser TAS instruments have become the main trend of neutron spectrometer instrumentation. The multi-analyser TAS concept originates from the so-called multiplexing analyser TAS such as RITA-II(Lefmann *et al.*, 2000) in the Paul Scherrer Institut. A multiplexing analyser is a design to split the analyser crystals into different analyser channels by implementing independent setting of crystal orientations. Due to the limited coverage of scattering angles, the multiplexing analyser TAS design was upgraded to the so-called multi-analyser design, in which the instrument covers a quite wide scattering angle (up to 70° or more) with many more analyser channels. The early example of a multi-analyser instrument is the FlatCone(Kempa *et al.*, 2006) at the Institut Laue–Langevin. Later, many other multi-analyser channel instruments were built, such as MultiFlexx, (Groitl *et al.*, 2017) MACS,(Rodriguez *et al.*, 2008) CAMEA,(Groitl *et al.*, 2016). These instruments could cover a wide scattering angle simultaneously with the same final energy. Some of them, such as CAMEA, could have several different final energies as well. That means one constant-Q scan can generate several different energy transfer results at a series of scattering channels. The data acquisition rate of such a TAS is significantly improved comparing to traditional TASs. Such a multi-final energy configuration is reminiscent of the multiplication repetition rate technique in the TOF spectrometers.(Mezei *et al.*, 2000) However, a challenge of such complicated TAS instruments is the data reduction and analysis. In order to help users for quickly and efficiently reduce their data collected on a multi-analyzer TAS, we introduce such a function in the TasData class.

In order to use the multi-analyser data reduction in TasData, the first step is to set up the configuration for the multi-analyser. The multi-analyser setup includes the number of analyser channels, the offset scattering angles of each channel comparing to the central channel, the detector group of each channel, and the efficiency of each single detector or detector wire in a position sensitive detector. With all these configuration settings, the function `tas_multichannel_reduction` in TasData calculates the actually scattering angle of each analyser channel, the positions of *qh*, *qk*, and *ql*, and the *en* values for each channel. It also corrects all the detector counts in the experiment with the corresponding data efficiency rate. Thus, the new *qh*, *qk*, *ql*, and *en* positions and the corresponding detector count can be grouped into one channel. Conducting all the similar calculations for all the channels, the data reduction of all the channels in the multi-analyser TAS complete. The results correspond to a group of points in the *qx*-*qy*-*en* space (*qx* and *qy* denote any combination of *qh*, *qk*, and *ql*) coordinate with the corresponding detector counts. TasData also provide a function *tas_multiana_plot3d* to visualize the reduced data set.

#### 3.1.4 Data Fitting with Convolution

Data fitting is one of the most important functions of *TasVisAn*. As mentioned above, the data plotting functions in both Taipan and Sika provide a quick peak fitting function using a Gaussian. It is also possible to fit a peak with a Lorentzian depending on the peak shape and the related physics. However, the most essential data fitting in TAS data analysis is to fit peaks in energy scans with a Lorentzian function by convoluting the instrument resolution at the current *Q* and *E* positions. *TasVisAn* implements these functions in both Taipan and Sika classes. With the `taipan_convfit` in Taipan and `sika_convfit` in Sika, one could fit the inelastic neutron peaks with instrument resolution convolution. Basically, these two functions used the instrument resolution calculation function and the convolution function in the TripleAxisSpectr class in the *InsPy* package to conduct the convolution calculation. The *fitter* class in *InsPy* can be used to conduct the fitting. Usually, the convolution fitting can be done in a *Jupyter notebook*, in which the fitting steps and the final fitted results will be shown. Alternatively, such convolution fittings can be conducted in the GUI program in the *InsPy* package without writing any code.

#### 3.1.5 Data Visualization

Both Taipan and Sika classes provide several easy solutions for plotting experimental data. For example, the `taipan_simpleplot` and `sika_simpleplot` functions can plot a single dataset or multiple datasets in one figure with the optional simple peak fitting function. The `taipan_combplot` and `sika_combplot` can simultaneously combine multiple datasets, normalize, and plot them in a single plot area with/without offset or a series of plot areas for comparison. Several other plotting functions are available as well. Most of these functions only need the users put the scan numbers as the input. One line of code is enough to plot the data users are interested in.

On the other hand, combining a series of scans into one intensity contour map is widely used for data visualization since it is more convenient to present phonon dispersion, spin wave dispersion, and diffuse scattering patterns in contour maps than in line plots. Both Sika and Taipan classes provide the contour plotting function. Users only need to arrange the scan numbers in a list and set up a few parameters like scanning axes, intensity scale. A contour plot can be plotted with the `tas_tidy_contour` and `tas_random_contour` functions. The data combination and normalization have been implemented in these functions. Users do not need to take care of these steps, which makes the data plotting much easier and convenient.

During the experiment, data could be collected in a random order, which makes it difficult to create contour plots using the two functions mentioned above. Alternatively, another two contour functions are provided to plot contour maps with data interpretation. Namely, these new functions will take the user data and interpret them over the whole area of the contour map. Users can add more and more datasets gradually following the progress of their experiments to improve the contour map with higher precision. These new functions can greatly help users to follow their experiment progress without effort. When sufficient data are collected, a contour map without interpretation can be plotted as the final version using the original contour functions mentioned earlier.

### 3.1.6 Scan Simulation & Validation

A TAS instrument most frequently operates in the reciprocal space by driving the virtual motors, such as *qh*, *qk*, *ql*, and *en*. Thus, users cannot directly know if these reciprocal positions are out of the accessible range of the real motors on the instrument. Additionally, a scan can be done only when the scattering triangle could be closed, namely, $k_i$, the indecent wavevector, $k_f$, the final wavevector, and Q, the momentum transfer, must form a triangle. Therefore, users should know if a scan could be conducted without driving motors out of their accessible range and make sure that the scattering triangle is closed. However, converting these virtual motor positions into the real motor positions is not that simple. To avoid the out-of-range and can-not-close-triangle issues during a TAS experiment, most TAS control software provides the function to calculate the real motor positions from the desired virtual motor positions, to assist users to make their decision how to perform their scans. Usually, however, this function is limited to the current configuration and the current scattering plane after the experiment has been setup on the TAS instrument. Namely, users could not use this function to plan their experiments in advance before their experiments start. Some TAS control software, such as the Labview-based Spice control software,(Lumsden *et al.*, 2006) creates a virtual instrument to allow users to setup their experiments and then conduct the scan simulation. However, users need to install software for the simulations, which is very complicated. *TasVisAn* provides the scan simulation functions `taipan_scansim` and `sika_scansim` in both Taipan and Sika classes with the default instrument configurations. Users can quickly check if their scans would go out of the limit of the motors of the spectrometers for their experiments. If they would like to change the setup, such as using a different $E_f$ for their experiments, they can conduct the simulation by simply changing the $E_f$ parameter in the *exp* parameter before starting the simulation. Examples will be shown in the later sections to demonstrate this function.

The modern design of TAS control software allows users to setup a long queue of scans. This requires users to be precise with their batch files. As well as the instrument accessible range problem discussed above, other easily avoidable problems are syntax mistakes and typographic errors in batch files. Any mistake in the batch files could cause instrument stops and beamtime loss. Thus, it is extremely important to validate and simulate the commands in the batch files. The Sika and Taipan classes in *TasVisAn* provide the functions to validate scan commands in batch files using the CommandValidator class in the *TasVisAn* package. The corresponding examples will be shown in the later section.

### 3.1.7 Experiment Log

Sika and Taipan classes also provide functions to export the experiment log file according to the collected data. Using the `export_scanlist` function in both classes, users can get a html report with the scan numbers and the corresponding descriptions of the scans in a table form. This function can significantly improve users' efficiency to track the experiment progress during the experiments and analyse the experiment data after the experiment.

### 3.1.8 Data Browser

A data browser with a GUI will be very useful for users to quickly go through the data. *TasVisAn* provides one integrated GUI data browser for both Taipan and Sika data. Figures 2 to 4 show the different pages of his browser. Users can choose the instrument at the top of the dialog for the data they would like to browser through. After choosing the experimental data folder, the browser will plot the first data in the first tab. As shown in Figure 2(a) and (b), Users can click the spin button in this data browsing tab to browser through all the data in this experiment. An automatic single peak fitting function can be enabled in this tab. They can check the data text (Figure 2(a)) and the reduced data (Figure 2(b)) in the text box below the plot area in this tab.

If users need to fit more than one peak, they can switch to the next "Graph and Fit" tab to do multiple peak fitting with control of the initial parameters. See Figure 3(a). This browser also allows users to plot contour map in the third "Contour" tab. Users can check all the scan information in the fourth "Scan History" tab (see Figure 3(b)), in which all the scans and scan titles are listed. All the scans are listed when users choose the data folder.

Users can conduct scan simulation for both Taipan and Sika in the last two tabs (see Figure 4 (a) and (b)) for experiment planning. By loading a batch file, users can validate all the scan commands in this batch file. Any syntax errors and typographic errors will be reported in the validation results. The total data points will be shown at the end of the report, which can help users to estimate the total time to complete all the scans in the batch file. After the batch file passes the validation, it can be submitted to the command queue of the instrument safely.

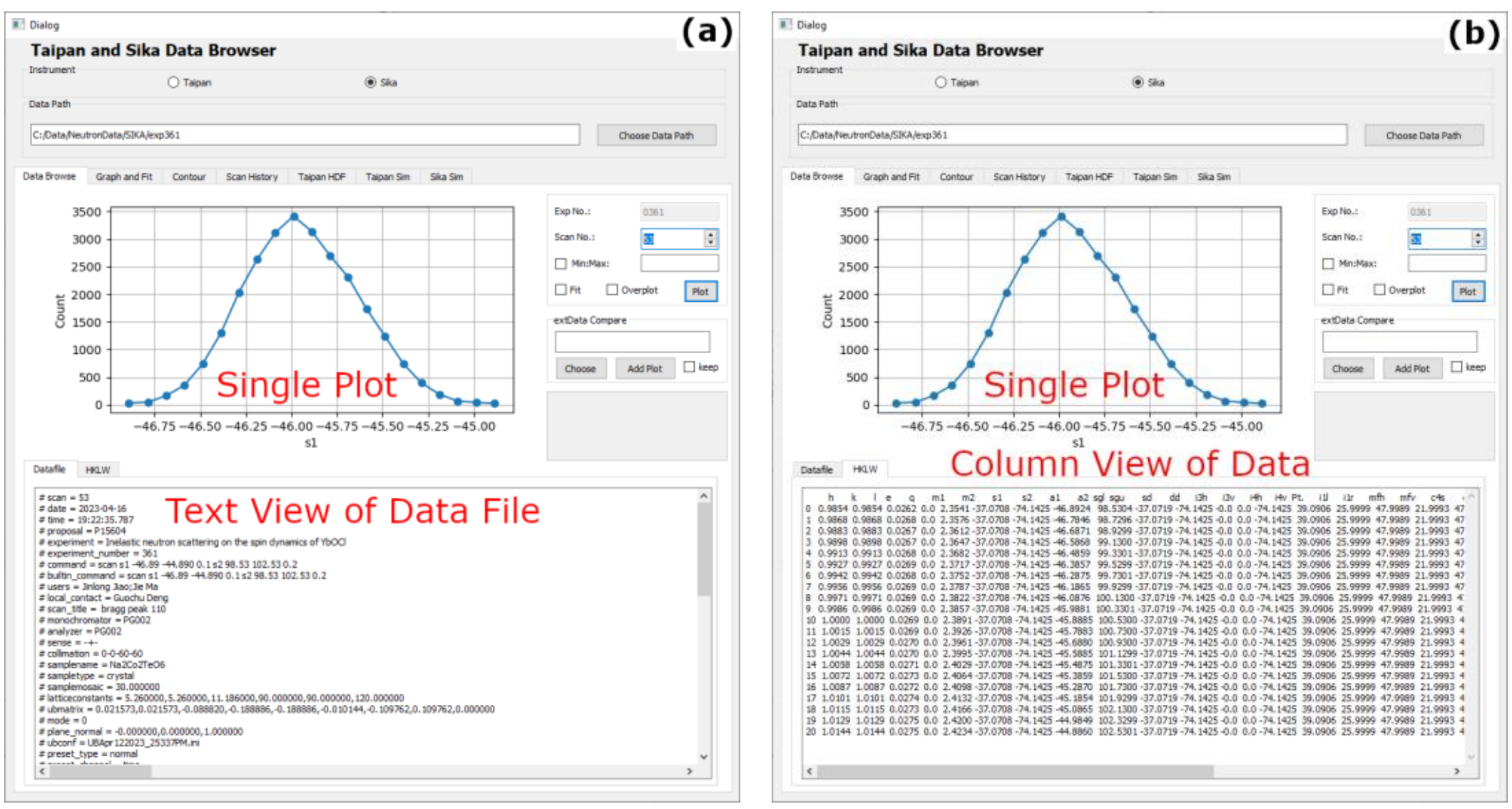

Figure 2 (a) TAS data browser with the original text data file (b) TAS data browser with the reduced column data

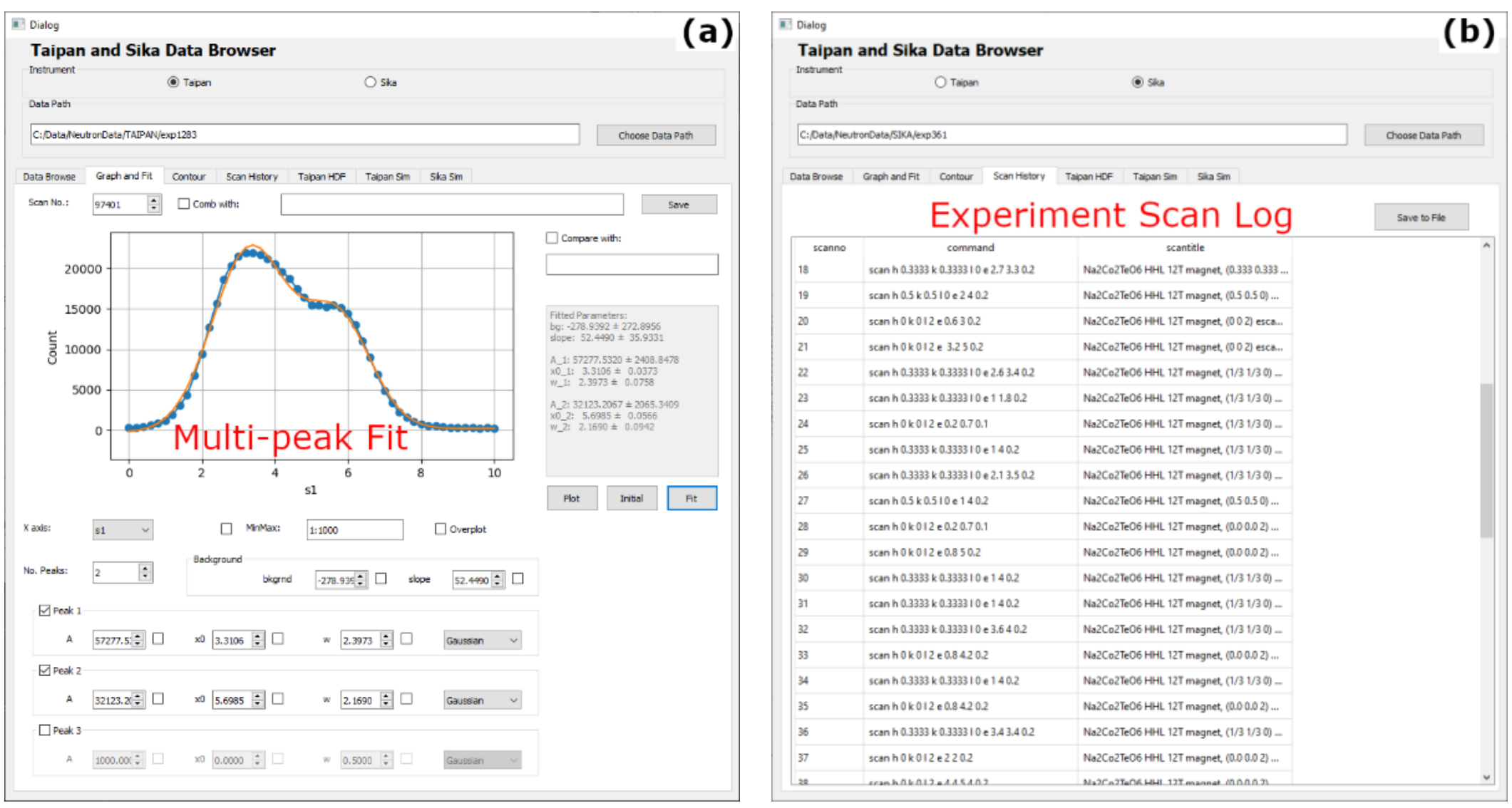

Figure 3 (a) Peak fitting tab of TAS data browser (b) Scan list summary tab of TAS data browser

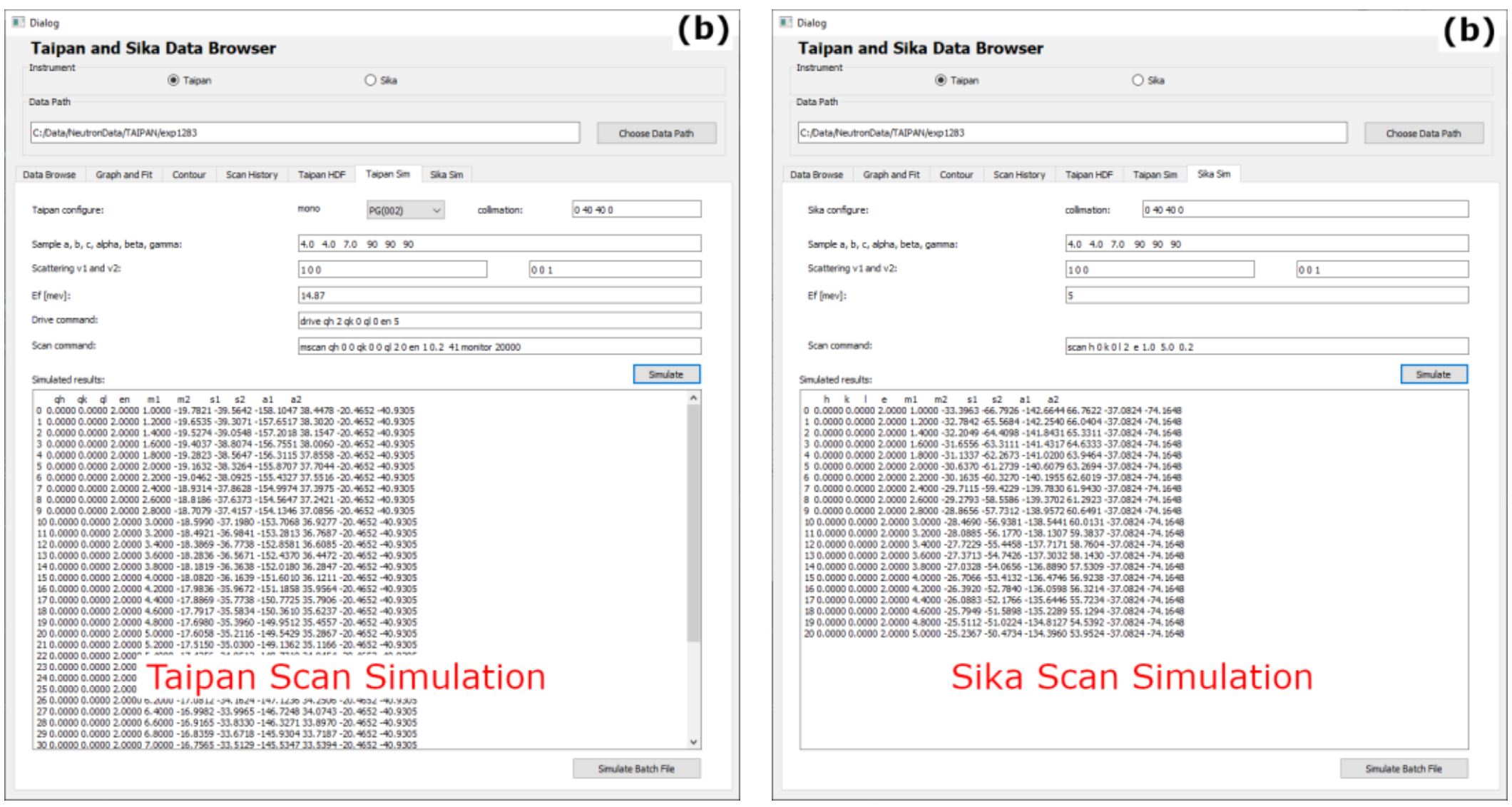

Figure 4 (a) Taipan scan simulation and batch file validation tab. (b) Sika scan simulation and batch file validation tab.

### 3.2 *InsPy* Package

*InsPy* is a Python package for calculating the instrument resolutions of TASs, which can be considered as a Python version of *Reslib*. It implements the two methods for the resolution calculation for a traditional TAS instrument, namely, the (Cooper & Nathans, 1967) and (Popovici, 1975) methods. With the instrument configurations such as the collimators, the arm length and the mosaicity of the crystals on the monochromator and analyser, and so on, *InsPy* can calculate the resolution ellipsoids in the 4D $Q$-$E$ space. *InsPy* also provides the data fitting functions by convoluting the instrument resolution.

#### 3.2.1 TAS Resolution Calculation

In general, the resolution of a TAS instrument varies at different $Q$ and $E$ positions. Considering the three-dimensional (3D) $Q$ and energy space, we should be able to use a 4D ellipsoid to describe the resolution function of a TAS instrument. Since we normally conduct an experiment in a scattering plane with energy changes, we only deal with a 3D resolution function like $R(Q_x, Q_y, E)$ in a real experiment.

The vertical divergence along $Q_z$ is normally not considered for the calculation. Experiments have also confirmed that the vertical focusing condition of the monochromator and analyser of a TAS does not impact its horizontal $Q$ and $E$ resolution significantly.(Shirane *et al.*, 2004) Taking the instrumental configurations, such as the dimension of the virtual source, the collimator divergences, the monochromator and analyser curvatures and the crystal mosaicities, the arm lengths, and so on, the instrument resolution can be calculated for any accessible ($Q$, $E$) positions analytically using the (Cooper & Nathans, 1967) method and the (Popovici, 1975) method.

Normally, the resolutions of a TAS instrument are plotted as the ellipses projected in the $Q_x$-$Q_y$ plane, the $Q_x$-E plane, and the $Q_y$-E plane. *InsPy* provides a function to plot the resolution functions in the three projected planes, as shown in Figure 5(a). Besides, *InsPy* also allows us to plot the resolution ellipsoids in the 3D $Q$-$E$ space. Figure 5 (b) shows a series of resolution ellipsoids of a TAS instrument at various energies in the $Q$-$E$ space, combining the 3D surface of a spin wave excitation spectrum.

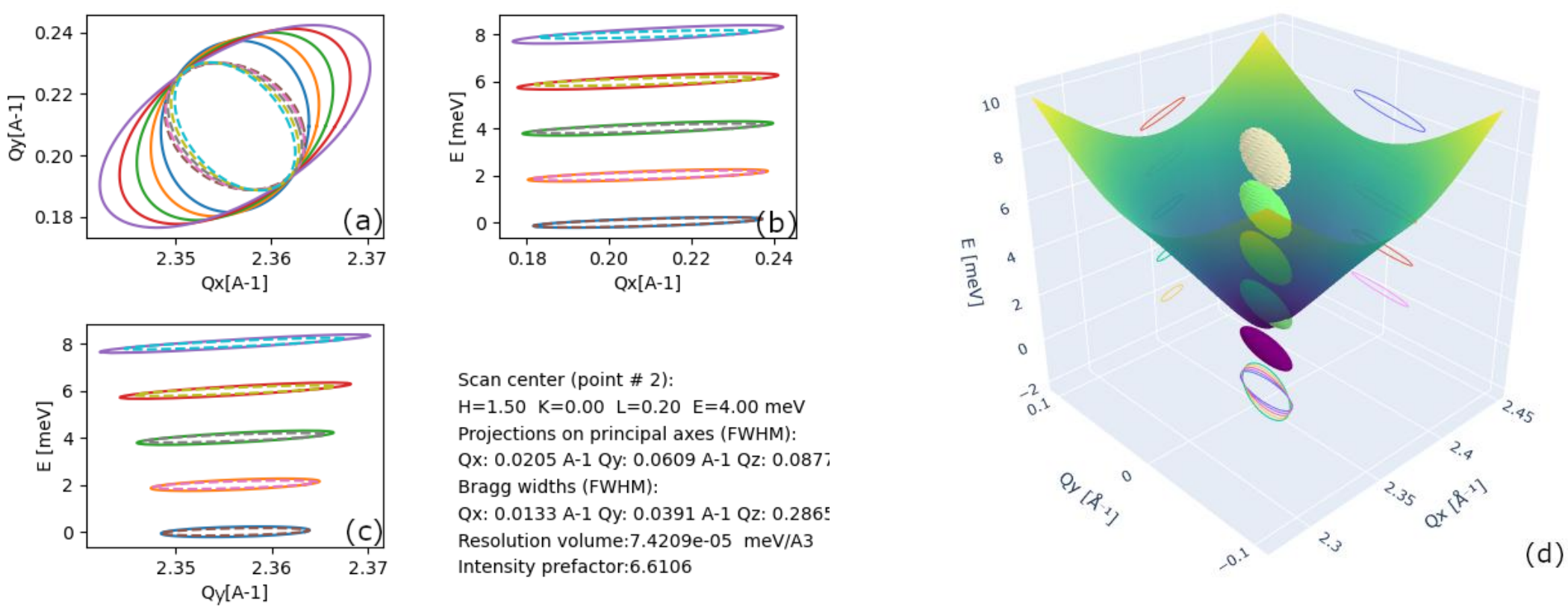

Figure 5 The projected resolution ellipses in the (a) $Q_x$-$Q_y$, (b) $Q_x$-$E$, and (c) $Q_y$-$E$ planes. (d) The 3D resolution ellipsoids of a TAS instrument in the $Q_x$-$Q_y$-$E$ space. The greenish curved surface shows the 3D spin wave dispersion in the $Q_x$-$Q_y$-$E$ space.

### 3.2.2 Data Fitting GUI

Calculating the resolution of a TAS instrument is not a simple task. However, *InsPy* provides an easy solution for this purpose. For the convenience of users without programming skills, the resolution calculation can be implemented in *InsPy* through the GUI program. As shown in Figure 6, *InsPy* provides a dialog for users to configure the instrument setup, calculate the instrument resolution, and conduct data fitting with the current setup. Users can setup the instrument configuration on the left panel. Any changes in this panel will be reflected in the instrument resolution in the middle panel. On the right panel, users can load their data and conduct data fitting with the instrument resolution calculated based on the configuration on the left panel. The program supports fitting two peaks simultaneously. However, users could manage to fit a single peak by properly setting the fitting parameters and fixing some of the parameters. The two GUI programs in Figure 6 (a) and (b) are designed to perform data fittings for energy scans and $Q$ scans, respectively. In both, the data range and the initial values of the parameters can be set by the users. Users can decide to fix any parameters in the fitting procedure. Depending on the type of data for fitting, users can enable or disable the magnetic form factor function by checking or unchecking the checkbox of “Mag Form Factor”. By typing the magnetic ion symbols in the input box on the upper-right corner in both (a) and (b), users can load the desired form factor for the data fitting. For energy scans, two Lorentzian functions are used to describe the two peaks according to the single mode Lorentzian approximation. While for $Q$ scans, two Gaussian functions are adopted to fit the two peaks according to the central limit theorem. For both energy and Q scans, the experiment temperature is used for the Bose factor correction. A user only needs to fix the intensity ratio to zero and fix all the parameters of the undesired peak for fitting the data with a single peak. The fitted parameter results and statistic information are shown on the screen and saved to simple text files.

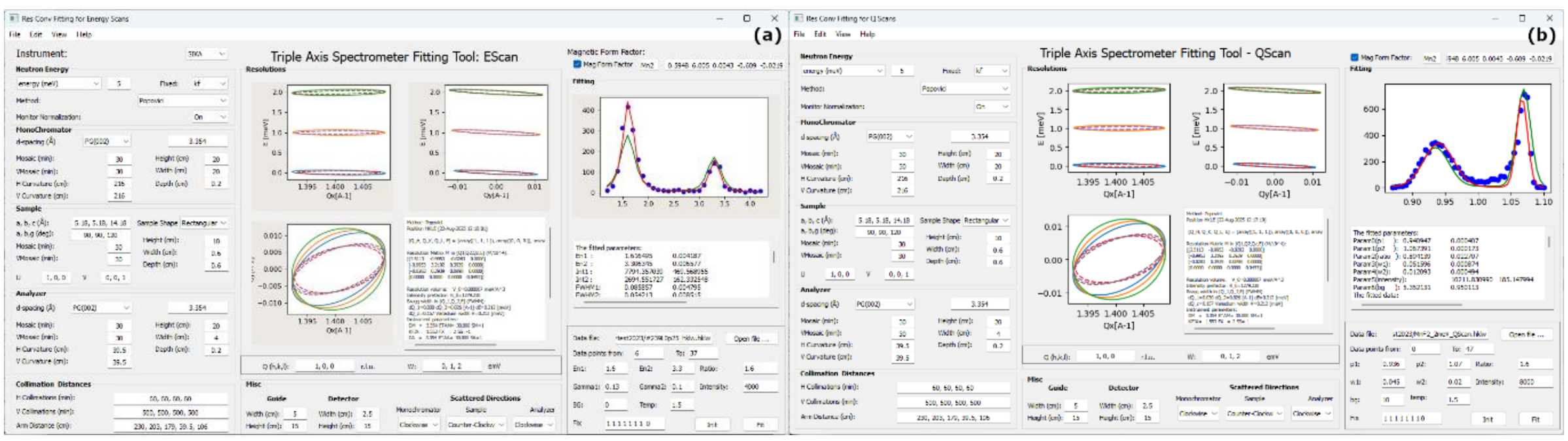

Figure 6 The *InsPy* GUI interfaces for calculating TAS resolutions, plotting resolution ellipses, and fitting data of (a) energy scans and (b) Q scans.

As demonstrated in Figure 6 (a), an energy scan with two excitation peaks was measured and plotted with the blue dots. The green curve shows the curve calculated with two Lorentzian peaks convoluted with the instrument resolution with the initial parameters given by users, while the red curve shows the fitted results through the resolution convolution fitting procedure described above. Figure 6 (b) shows a *Q* scan (or a constant-energy scan) with double excitation peaks in a blue dotted curve, corresponding to the left and right branches of the spin wave excitation. Apparently, the curve shows a strong focusing effect with a broad peak on the left and a very sharp on the right. The green curve is a curve calculated from two Gaussian peaks convoluted with the instrument resolution. The red peak shows the final fitted result using the convolution fitting method.

### 3.2.3 Asymmetric peak fitting

As mentioned above, energy scans from a TAS instrument are usually fitted by using a Lorentzian convoluted with the instrument resolution.(Zheludev, 2009) Another widely used model to fit inelastic neutron excitation peak is the so-called damped harmonic oscillator (DHO) model.(Hansen *et al.*, 1997) Both the Lorentzian function and the DHO model function are symmetric peak functions. These symmetric functions are included in the *InsPy* for the data fitting. However, asymmetric excitation peaks are widely observed in inelastic neutron scattering experiments, especially in measuring magnetic excitations such as spin wave excitations. Thus, we need to introduce some asymmetric peak functions to fit those asymmetric peaks observed in TAS experiments.

In *Inspy*, two asymmetric peak functions are available for users to conduct their fittings to asymmetric peaks. The first asymmetric peak function is based on the dynamical susceptibility function proposed by Mason *et al.*, (1992) and Raymond *et al.*, (1997). It has the following form:

$$\chi'' = A|\varepsilon|\omega\left[Re\left(1/\sqrt{\omega^2-\varepsilon^2}\right)\right]^2 \tag{3}$$

where $\varepsilon = \Delta + i\,\Gamma$, in which $\Delta$ is inelasticity, and $\Gamma$ is the linewidth; ω is the energy. This asymmetric function was originally proposed to describe excitation in Kondo metals and seems to be applicable in other scenarios.(Mason *et al.*, 1992)

Even though the DHO model above is widely used in fitting inelastic neutron scattering data, the limits of DHO model are clear, as discussed above. Due to its slow decay at high energy, the DHO model usually cannot fit data well in a broad energy range. Here, we consider building a modified DHO model. In this new model, we assume that the potential of the system is not a parabola by adding an extra term in the recovery force $F = -k_0x - k_1x^2$, namely, an extra cubic term in the potential $V(x)$. Thus, the potential becomes asymmetric, and we call it a damped anharmonic oscillator (DAO) model.

Following the similar steps of deriving the DHO model, we introduce the damping term $i\Gamma(\omega)$ and add the frequency shift term $\Delta(\omega)$. Thus, we get the effective energy as follows:

$$\epsilon_{\mathrm{eff}}^2(\omega) \approx \left[\omega_0 \ + \ [\Delta(\omega) + i\Gamma(\omega)]\right]^2$$

where $\epsilon_{\mathrm{eff}}$ is the effective energy, $\Delta(\omega)$ is the frequency-dependent energy shift, and $\Gamma(\omega)$ the frequency-dependent damping term. We obtain a phenomenological damped anharmonic scattering function:

$$I(Q,\omega) = I_{BG} + [n(\omega)+1]\omega A_Q \left[\mathrm{Re}\frac{1}{\sqrt{\omega^2 - [\omega_0 + \Delta(\omega) + i\Gamma(\omega)]^2}}\right]^2$$

Where, $I_{BG}$ is the background, $[n(\omega)+1]$ is the Bose factor, $A_Q$ is the amplitude, the denominator is the revised one as described above. In the formula, $\Delta(\omega) = \sigma(\omega - \omega_0)$, which suggests that the energy shift scales linearly with the distance from the resonance, reflecting the linear change in Hooke's coefficient $k(x)$. The damping term in the formula is not constant but a function of the energy, namely, $\Gamma(\omega) = \Gamma^2 + \alpha(\omega - \omega_0)^2$ , indicating that the damping becomes stronger as pushing the system into the anharmonic regime.

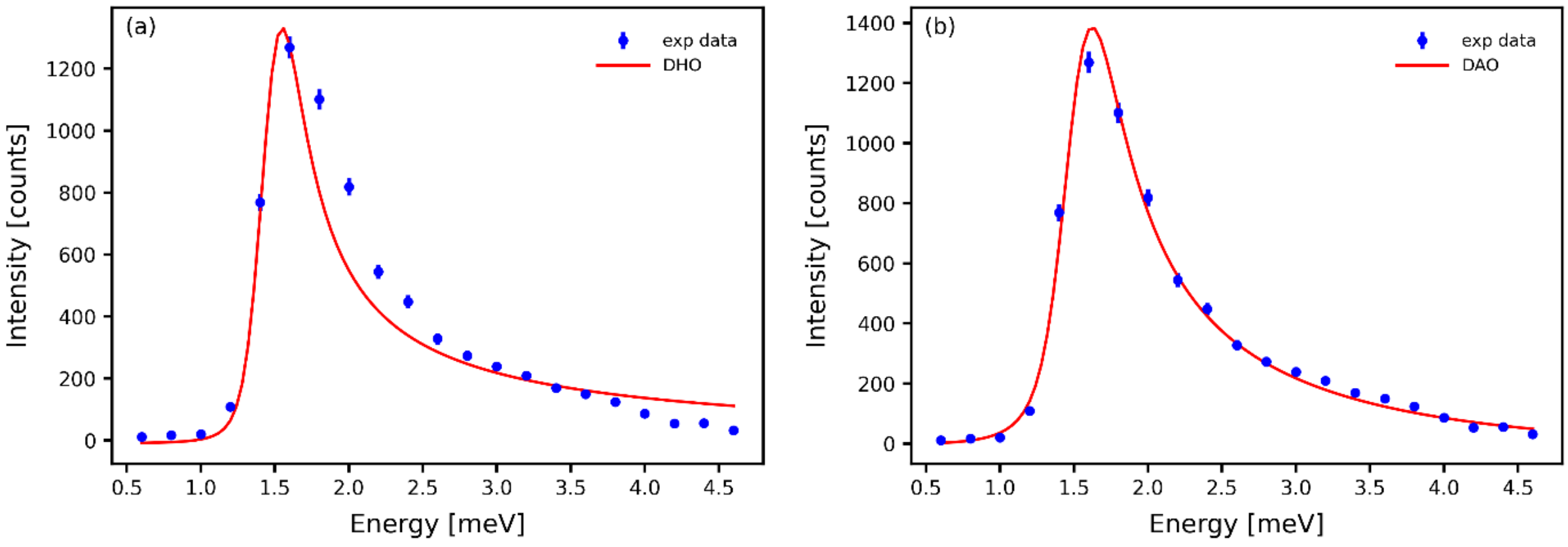

Figure 7 (a) Fitting an energy scan of spin wave excitation near its zone centre with the first asymmetric function discussed in the text. (b) Fitting the same energy scan with the asymmetric Lorentzian function discussed in the text.

To confirm if the new asymmetric peak function described above is suitable for data fitting, we compare the final fitting results from these two functions when fitting to the same dataset. The data was a spin wave excitation peak collected from a single crystal of $Mn_4Nb_2O_9$ on Sika at ACNS.(Deng *et al.*, 2022) The fitted results are shown in Figure 7, in which (a) is from the first asymmetric function and (b) is from the new DAO asymmetric function. Figure 7 clearly shows that the second asymmetric function fits the experimental data much better than the first function. The peak centre in the DAO fit looks more consistent with the data. The DAO fitted result produces an intensity decay profile in Figure 7 (b) which highly agree with the experimental data, much better than the DHO model in Figure 7 (a). A more systematically comparison study about these two functions and another two empirical functions was presented in our previous paper. (Deng, 2025)

## 4. Usage Examples

### 4.1 Simple Curve Plotting and Fitting

Quickly visualizing data from a TAS instrument is so important for users to make decision about the next step during the experiment. One of the main functions of the *TasVisAn* package is to help users to plot their experimental data easily during their experiment. The code in Figure 8 demonstrates how to plot several line plots in one graph with the simple Gaussian fitting and an offset setting. Basically, one line Python code is sufficient for plotting the data just like Figure 9. With the fit option to be true, all the peaks will be fitted, and the fitted parameters will be returned. Thus, the comparison between these four peaks will be so clear with the output of the fitted parameters, such as, peak amplitudes, peak widths, etc. If turning overplot to false, each curve will be plotted in a separated panel in case users prefer a different view.

```
import matplotlib.pyplot as plt
from acnstas.Sika import Sika

sika= Sika( expnum="155", title="SpinWave MNO", sample="MNO", user=['G. Deng et al.'])
path='Datafiles/'
scanlist =   [ 180, 179, 178, 176]
pars,fitdata,fig, ax=sika.sika_combplot(path, scanlist, fit=True, overplot=True, offset=1000)
customize_axis(ax, "Diffuse Scattering Near T$_N$", xlabel="Q$_H$ [r.l.u.]", ylabel="Intensity [a.u.]")
```

Figure 8 Python code in *Jupyter notebook* to demonstrate how to plot multiple line curves from Sika data.

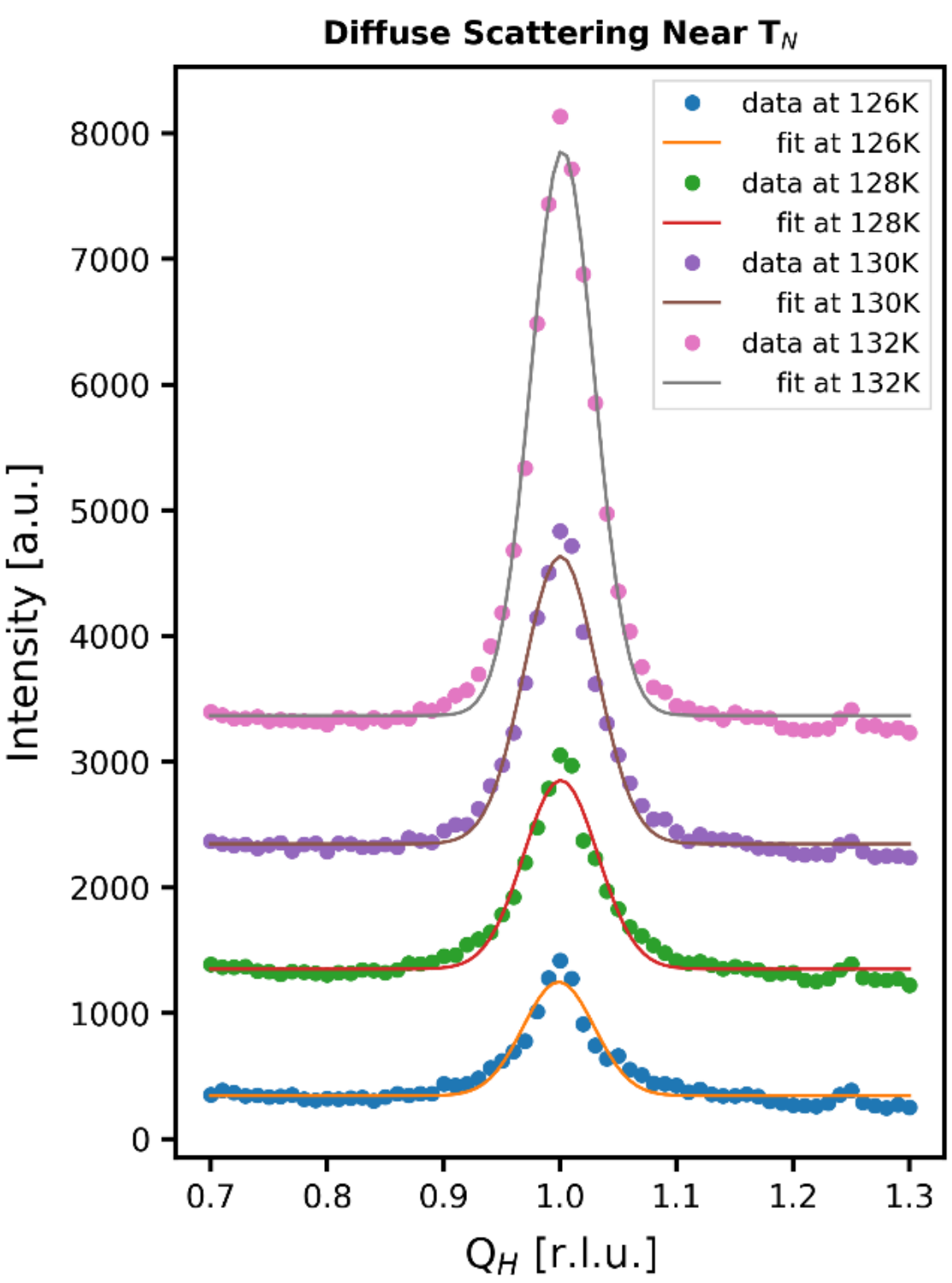

Figure 9 The line plot generated by the code in Figure 8. The curves in this plot were fitted to Gaussians with an offset along the vertical axis by 1000 between any two neighbouring curves.

### 4.2 Data Concatenation by Row

In a TAS experiment, many users would like to collect peak intensities of a structural or magnetic Bragg peak at different temperature. This can provide the Bragg peak intensity as a function of temperature, which is very useful for analysing the phase transition dimensionality. Considering the time efficiency, a so-called sit-and-count technique is widely used for collecting temperature-dependent data. Basically, we could drive the instrument to the Bragg peak centre and keep the instrument in a static status and slowly change the sample temperature upon heating. Simultaneously, the instrument counts the intensity and record the results in a series of data files. With precisely controlled heating speed and sufficient counting statistics, beautiful temperature dependences of the Bragg intensity can be collected. Usually, hundreds of data files will be collected in this procedure, which is hard to be handled manually. Both Taipan and Sika classes provide a function to quickly plot those temperature-dependent curves. Figure 10 presents some example code to import the temperature and counts from more than two hundred data files. With some further polishing code, a publishable figure can be generated from the data, as demonstrated in Figure 11.

```
sika          = Sika( expnum="304", title="NCSO Under Magnetic Fields", sample="NCSO", user=['Sika User'])
path          = 'Datafiles/'
scanlist      = np.arange(100, 500, 1)
data, _       = sika.data_red_anycolumn(path, scanlist=scanlist, columnlist=['tc1a','tc2a', 'count'], mode='byrow')
sorted_data   = data.sort_values(by=['tc1a'])
sorted_data   = sorted_data[sorted_data['tc1a'] >1.0]
plt.plot(sorted_data['tc1a'], sorted_data['count'])
```

Figure 10 Python code in *Jupyter notebook* to demonstrate how to combine many data files into one temperature-dependent peak intensity.

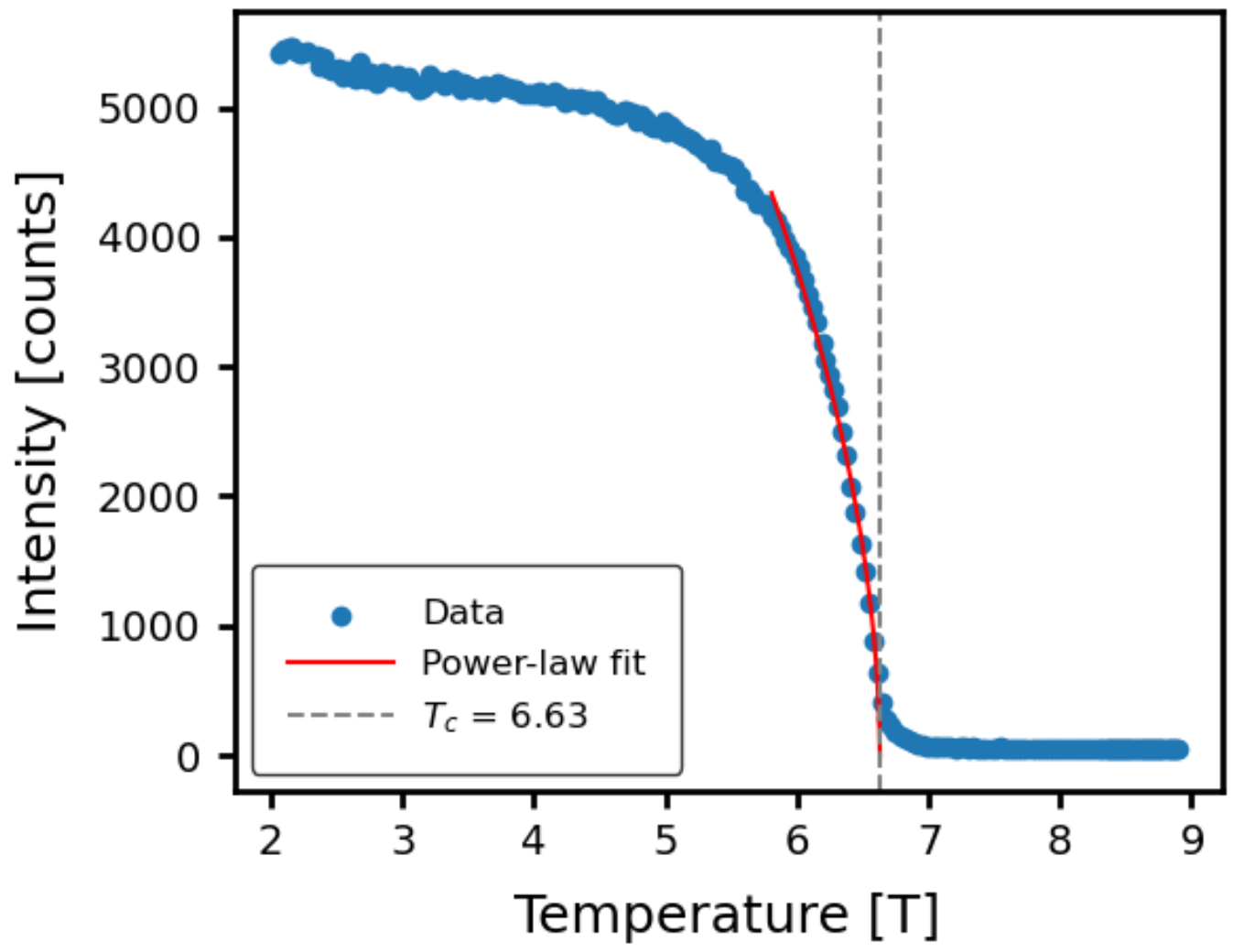

Figure 11 Magnetic Bragg peak intensity as a function of temperature measured using the "sit-and-count" method. The data was fitted to the power-law function in the critical temperature range.

#### 4.3 Data Contour Map

It is common to conduct a series of scans to map spin wave dispersion, phonon dispersion, diffuse scattering, etc. of the sample during a TAS experiment. There is no easy way to plot all these scans together to create a contour map by handling the data manually. To solve this problem, we implement the functions `tas_tidy_contour` in TasData, respectively, to help users to generate contour maps without effort. The code in Figure 12 demonstrates how to create a contour map of phonon dispersion with the data collected on Sika. Figure 13 shows a generated contour map from the phonon dispersion measured in a $Hg_2Br_2$ single crystal.(Zhu *et al.*, 2019)  The phonon dispersion along this direction is clearly shown. This function is highly useful to help users to check if all the necessary data are collected and sufficient statistics achieved during an experiment.

```
import pandas as pd
from tasvisan.tas import Sika
sika= Sika( expnum="270", title="phonon", sample="Hg2Br2", user=['A. B', 'C. D'])
path='Datafiles/'
scanlist =  [153, 152, 154, 151, 155, 150, 156, 149, 157, 147]
motorlist = ['qh', 'qk', 'ql', 'en', 'detector', 'monitor']

dflist=sika.sika_batch_reduction(path, scanlist, motorlist=motorlist)
qq, ee, intt, ax =sika.tas_tidy_contour(dflist,  x_col='qk', y_col='en', vminmax=[20,3000])
ax.set_xlabel("qK [r.l.u.]",fontsize=18, labelpad=10)
ax.set_ylabel("Energy Transfer [meV]",fontsize=18, labelpad=10)
ax.tick_params(axis='both', which='major', labelsize=14, width=2, length=6, pad=5)
ax.tick_params(axis='both', which='minor', width=1.5, length=4)
for spine in ax.spines.values():
    spine.set_linewidth(2)
```

Figure 12 Python code demonstration in *Jupyter notebook* to plot the intensity contour map of the phonon dispersion along the $Q_H$ direction with the Sika class in the *TasVisAn* package

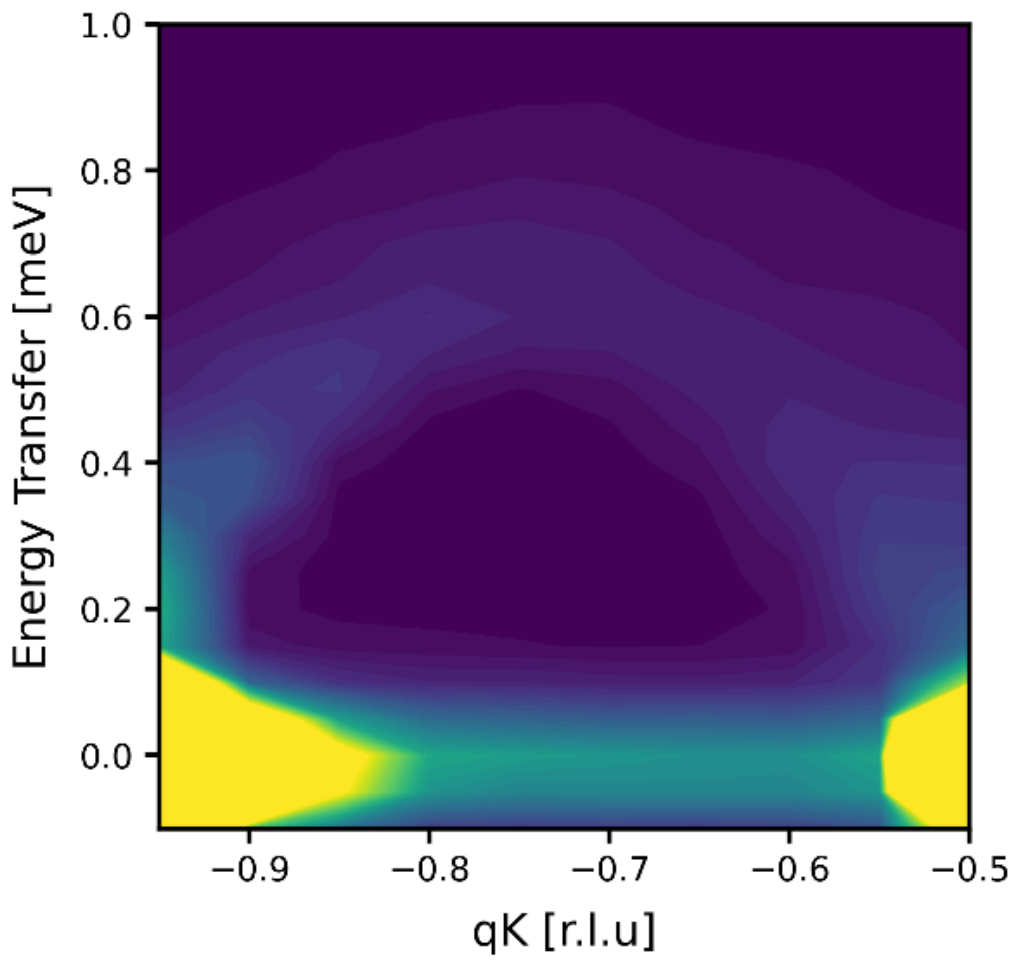

Figure 13 The intensity contour map generated using the code in Figure 12 by importing the Sika data from $Hg_2Br_2$.

### 4.3 Visualization of Multi-analyser Data in 3D

Figure 14 illustrates the code used to configure the multi-analyser setup for the *multi-Q constant-$E_f$* multiplexing mode on Sika, as well as to perform data reduction for measurements collected in this mode. The data shown were acquired from a phonon dispersion experiment on a lead single crystal, with Sika operating in the *multi-Q constant-$E_f$* configuration. In this mode, all analyser blades are set to the same final energy ($E_f$), but each blade corresponds to a unique scattering angle, thereby sampling a different momentum transfer ($\boldsymbol{Q}$). To carry out data reduction across all analyser channels, the configuration must specify the scattering angle offsets for each channel, apply detector efficiency corrections, and define the detectors associated with each analyser. The reduced multiplexed data can then be visualized in three dimensions using the mplot3d module from the *matplotlib* library, as shown in Figure 15. This interactive 3D visualization allows users to freely rotate and explore the data from different perspectives.

```
from tasvisan.tas import Sika

sika= Sika( expnum="24", title="Multiplexing", sample="Pb", user=['Sika User'])
path='Datafiles/'
sika.sika_multichannel_config()

scanlist=[492, 494, 495, 496]
alldflist=[]
for scan in scanlist:
    df = sika.sika_data_to_pd(path, scan)
    dflist = sika.tas_multichannel_reduction(df)
    alldflist = alldflist+dflist

sika.tas_multichannel_plot3d(alldflist, xcol='qh', ycol='qk', zcol="en", perc_low=5, perc_high=95,
            xlim=None, ylim=None, zlim=None,xlabel="Q$_H$ [r.l.u.]", ylabel="Q$_K$ [r.l.u.]",zlabel="E [meV]")
```

Figure 14 Python code demonstrating how to configure the multiplexing operational mode on Sika and reduce the multiplexed data.

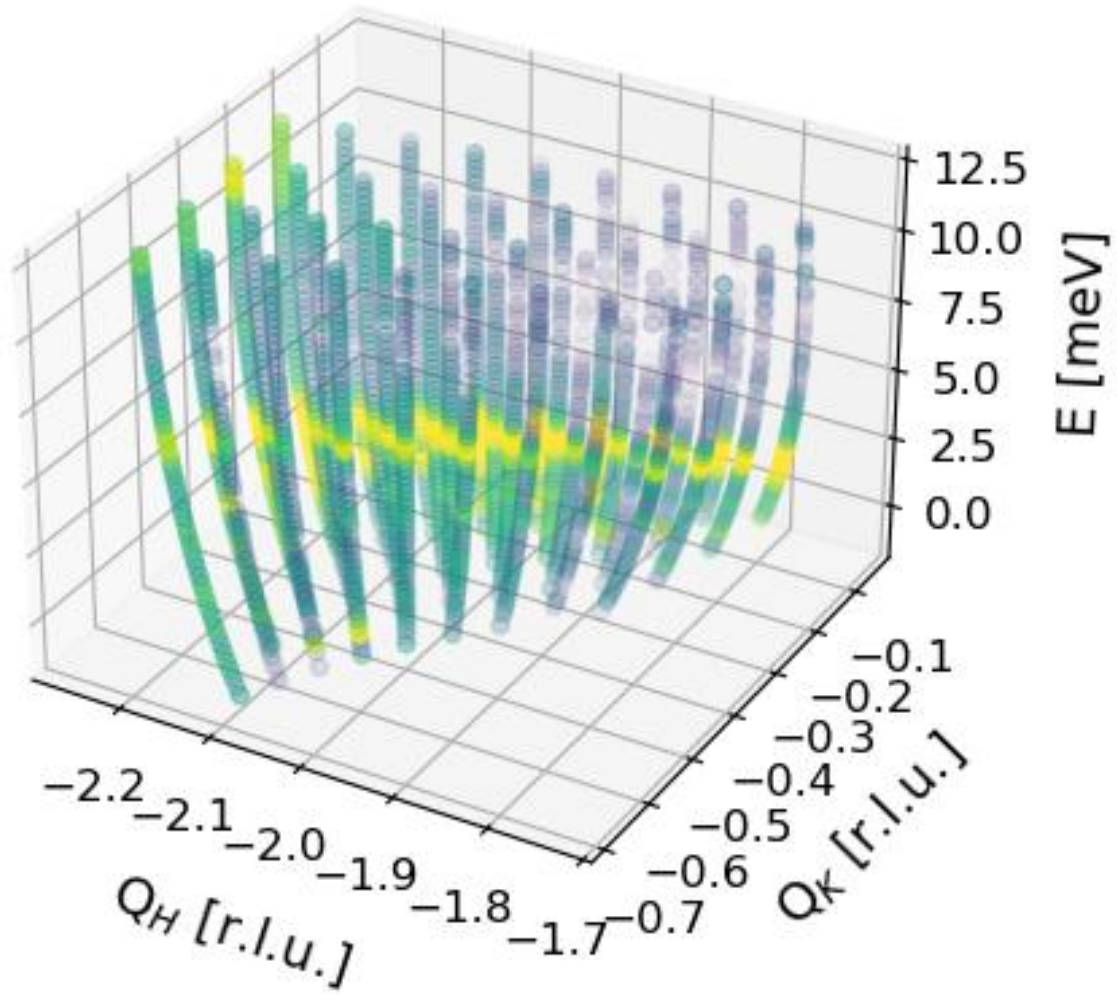

Figure 15 The visualization of the 3D multiplexed data collected on Sika and reduced using the code shown in Figure 14. The data was collected from a piece of lead single crystal sample with nine multiplexing channels in use. The scattering angle and the detector data efficiency were determined by conducting a diffraction experiment.

#### 4.4 Scan Simulation and Scan Batch Validation

As mentioned above, the Taipan and Sika classes provide the scan simulation functions. Figure 16 (a) and (b) shows the scan simulations on Taipan and Sika, respectively. The program can give warnings to users when there are problems in the proposed scan, such as syntax errors or geometric issues. For example, both these two simulations shown the warning of "scattering triangle cannot close". The simulated results also provide the angular information of each motor for users to assess if these angles are in a proper range. They can avoid wasting time to conduct scans in the range with high background, e.g. in the low scattering angle range. This scan simulation function is very useful for users to plan their scans and run their experiment efficiently.

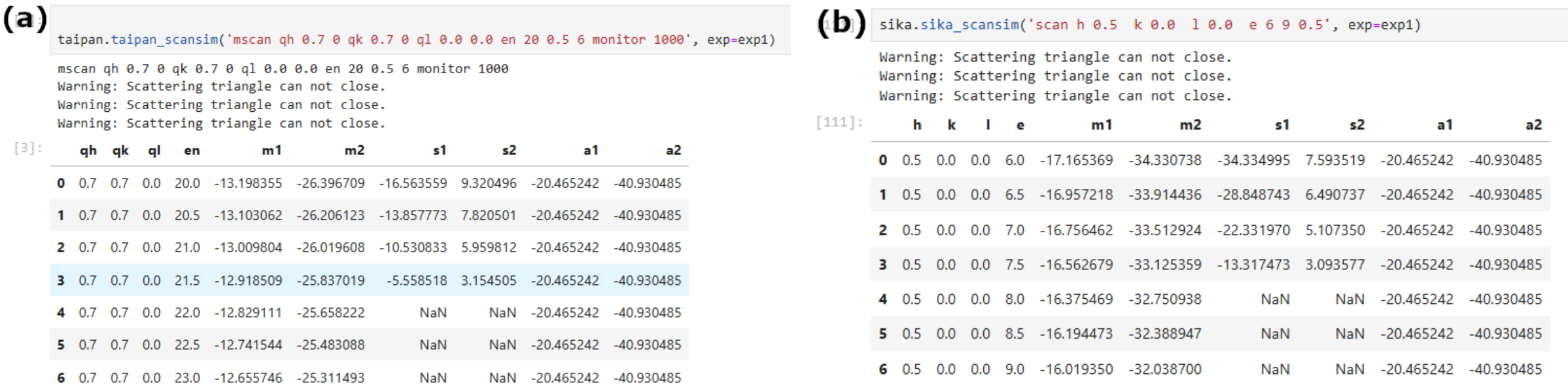

(a)

```
taipan.taipan_scansim('mscan qh 0.7 0 qk 0.7 0 ql 0.0 0.0 en 20 0.5 6 monitor 1000', exp=exp1)
```

```
mscan qh 0.7 0 qk 0.7 0 ql 0.0 0.0 en 20 0.5 6 monitor 1000
Warning: Scattering triangle can not close.
Warning: Scattering triangle can not close.
Warning: Scattering triangle can not close.
```

[3]:

| | qh | qk | ql | en | m1 | m2 | s1 | s2 | a1 | a2 |
|---|---|---|---|---|---|---|---|---|---|---|
| 0 | 0.7 | 0.7 | 0.0 | 20.0 | -13.198355 | -26.396709 | -16.563559 | 9.320496 | -20.465242 | -40.930485 |
| 1 | 0.7 | 0.7 | 0.0 | 20.5 | -13.103062 | -26.206123 | -13.857773 | 7.820501 | -20.465242 | -40.930485 |
| 2 | 0.7 | 0.7 | 0.0 | 21.0 | -13.009804 | -26.019608 | -10.530833 | 5.959812 | -20.465242 | -40.930485 |
| 3 | 0.7 | 0.7 | 0.0 | 21.5 | -12.918509 | -25.837019 | -5.558518 | 3.154505 | -20.465242 | -40.930485 |
| 4 | 0.7 | 0.7 | 0.0 | 22.0 | -12.829111 | -25.658222 | NaN | NaN | -20.465242 | -40.930485 |
| 5 | 0.7 | 0.7 | 0.0 | 22.5 | -12.741544 | -25.483088 | NaN | NaN | -20.465242 | -40.930485 |
| 6 | 0.7 | 0.7 | 0.0 | 23.0 | -12.655746 | -25.311493 | NaN | NaN | -20.465242 | -40.930485 |

(b)

```
sika.sika_scansim('scan h 0.5  k 0.0  l 0.0  e 6 9 0.5', exp=exp1)
```

```
Warning: Scattering triangle can not close.
Warning: Scattering triangle can not close.
Warning: Scattering triangle can not close.
```

[111]:

| | h | k | l | e | m1 | m2 | s1 | s2 | a1 | a2 |
|---|---|---|---|---|---|---|---|---|---|---|
| 0 | 0.5 | 0.0 | 0.0 | 6.0 | -17.165369 | -34.330738 | -34.334995 | 7.593519 | -20.465242 | -40.930485 |
| 1 | 0.5 | 0.0 | 0.0 | 6.5 | -16.957218 | -33.914436 | -28.848743 | 6.490737 | -20.465242 | -40.930485 |
| 2 | 0.5 | 0.0 | 0.0 | 7.0 | -16.756462 | -33.512924 | -22.331970 | 5.107350 | -20.465242 | -40.930485 |
| 3 | 0.5 | 0.0 | 0.0 | 7.5 | -16.562679 | -33.125359 | -13.317473 | 3.093577 | -20.465242 | -40.930485 |
| 4 | 0.5 | 0.0 | 0.0 | 8.0 | -16.375469 | -32.750938 | NaN | NaN | -20.465242 | -40.930485 |
| 5 | 0.5 | 0.0 | 0.0 | 8.5 | -16.194473 | -32.388947 | NaN | NaN | -20.465242 | -40.930485 |
| 6 | 0.5 | 0.0 | 0.0 | 9.0 | -16.019350 | -32.038700 | NaN | NaN | -20.465242 | -40.930485 |

Figure 16 (a) Demonstration of the scan simulation function of Taipan (b) Demonstration of the scan simulation function of Sika

#### 5. Discussion and Future Work

The current version of *TasVisAn* implement most of the common functions for data reduction, visualization and fitting for both traditional TAS instruments and modern multi-analyser systems. One of the primary tasks for future development is to add more classes to the package that are compatible with TAS instruments from other facilities. With these upgrades, *TasVisAn* will enable a broader user base to analyse data collected on various TAS instruments more efficiently.

In addition, some multi-analyser TAS instruments such as CAMEA(Groitl *et al.*, 2016) have different configurations, which is not yet implemented in the current version. In the future, we will include data reduction and visualization of the data from varieties of TAS instruments with special designs. Therefore, it is important to extend the advanced data analysis capabilities of the package to support many more TAS systems. Following the trend of more and more 2D data available in many newly built TASs, we also consider providing functions which could directly fit 2D maps with instrumental resolution convolution in the future.

Furthermore, as the number of time-of-flight (TOF) spectrometers increases globally,(Copley & Udovic, 1993) it becomes important to implement resolution function calculations for a variety of TOF instruments. However, since TOF instrument designs differ significantly between pulsed and continuous sources, there is no general method for resolution calculation that applies to all of them. As such, this feature will be implemented in *InsPy* on an instrument-by-instrument basis.

## 6. Conclusion

We have implemented two Python packages, *TasVisAn* and *InsPy*, for data analysis and resolution calculation on TAS instruments. *TasVisAn* provides functions for data reduction, visualization, and fitting, while *InsPy* enables users to compute instrument resolution and perform resolution-convolved data fitting. Several GUI tools have been implemented to assist users without programming experience in browsing data and conducting fittings. Examples are provided to demonstrate the capabilities of both packages.

## Supporting Information

The files provided in the supporting information for the article are as follows:

(i) TasVisAn_Demo.html – the exported html format of Jupyter notebook to demonstrate how to use TasVisAn.

(ii) TasVisAn_Demo.ipynb – the Jupyter notebook example to demonstrate how to use TasVisAn.

(iii) Taipan_expdemo – the raw Taipan experiment data for the Jupyter notebook demonstration.

(iv) TasVisAn_GUI_demo.mp4 – the screenshot video to demonstrate how to use the TasVisAn TASDataBrowser to browse the experiment data and simulate scans.

(v) Inspy_GUI_demoA.mp4 – the first screenshot video to demonstrate how to use the Inspy GUIs to calculate the instrument resolution and fit the experiments by convoluting with the instrument resolution.

(vi) Inspy_GUI_demoB.mp4 – the second screenshot video.

## Acknowledgements

The authors gratefully acknowledge the support of Australian Centre for Neutron Scattering (ACNS) at Australian Nuclear Science and Technology Organization (ANSTO). The authors also thank ACNS for the allocation of neutron beamtime for the data used for demonstration in this work.

## Conflicts of interest

The authors declare no conflicts of interest.

#### Data availability

Data in this article is available upon request.